\documentclass[12pt,preprint]{aastex}

\usepackage{natbib}
\usepackage{enumerate}
\usepackage{enumitem}

\slugcomment{}

\shorttitle{Confined Flares in AR 12192}
\shortauthors{Chen et al.}

\begin{document}


%\title{EUV Arch preventing eruption of a filament , \\
%    Gauge-Boson Couplings, and AAS\TeX\ Examples}
%\title{Overlying Extreme-ultraviolet Arcades Preventing Eruption of a
%        Filament Observed by AIA/SDO}
%\title{Direct Observations of Tether-cutting Reconnection During a Major
 %       Solar Event on 2014 February 25}
\title{Confined Flares in Solar Active Region 12192 from 2014 October 18 to 29}
%\title{EUV Observations of T-C reconnection During a Major Solar Event by SDO/AIA}

\author{Huadong Chen\altaffilmark{1}, Jun Zhang\altaffilmark{1}, Suli Ma\altaffilmark{2}, 
Shuhong Yang\altaffilmark{1}, Leping Li\altaffilmark{1}, Xin Huang\altaffilmark{1}, Junmin Xiao\altaffilmark{1}}
\email{hdchen@nao.cas.cn}
%\affil{College of Science, China University of Petroleum, Qingdao 266580, China}
%\affil{Key Laboratory of Solar Activity, National Astronomical Observatories, Chinese Academy of Sciences, Beijing 100012, China}
\altaffiltext{1}{Key Laboratory of Solar Activity,
     National Astronomical Observatories, Chinese Academy of Sciences,
      Beijing 100012, China}
\altaffiltext{2}{College of Science, China University of Petroleum, Qingdao 266580, China}
%\author{S. Djorgovski\altaffilmark{1,2,3} and Ivan R. King\altaffilmark{1}}
%\affil{Astronomy Department, University of California,
%    Berkeley, CA 94720}

%\author{C. D. Biemesderfer\altaffilmark{4,5}}
%\affil{National Optical Astronomy Observatories, Tucson, AZ 85719}
%\email{aastex-help@aas.org}

%\and

%\author{R. J. Hanisch\altaffilmark{5}}
%\affil{Space Telescope Science Institute, Baltimore, MD 21218}

%\altaffiltext{2}{Society of Fellows, Harvard University.}
%\altaffiltext{3}{present address: Center for Astrophysics,
%    60 Garden Street, Cambridge, MA 02138}
%\altaffiltext{4}{Visiting Programmer, Space Telescope Science Institute}
%\altaffiltext{5}{Patron, Alonso's Bar and Grill}


\begin{abstract}
Using the observations from the Atmospheric Imaging Assembly (AIA) and Helioseismic and Magnetic Imager (HMI) aboard the $Solar\ Dynamics\ Observatory$ ($SDO$), we investigate six X-class and twenty-nine M-class flares occurring in solar active region (AR) 12192 from October 18 to 29. 
\textbf{Among them, thirty} (including six X- and twenty-four M-class) flares originated from the AR core \textbf{and the other five M-flares appeared at the AR periphery.}
Four \textbf{of the} X-flares exhibited similar flaring structures, indicating they were homologous flares with analogous triggering mechanism.
%According to the $SDO$ observations, 
The possible scenario is: photospheric motions of emerged magnetic fluxes lead to shearing of the associated coronal magnetic field, which then yields a tether-cutting favorable configuration.
%According to the AIA observations, it is likely that tether-cutting reconnections occurred between the sheared arcades during these homologous flares, which may have a close association with .
\textbf{Among the five periphery M-flares, four} were associated with jet activities.
%Four of the five M-flares from the AR periphery were associated with jet activities.
The HMI vertical magnetic field data show that the photospheric fluxes of opposite magnetic polarities emerged, converged and canceled with each other at the footpoints of the jets before the flares.
Only one M-flare from the AR periphery was followed by a coronal mass ejection (CME).
%From October 20 to 26, \textbf{the total average} of the mean decay indices $\overline{n}$ of the horizontal background fields \textbf{is 1.40$\pm$0.13}, 
From October 20 to 26, the mean decay \textbf{index} of the horizontal background field within the height range of 40--105 Mm \textbf{is} below the typical threshold for torus instability onset. This suggests that a strong confinement from the overlying magnetic field might be responsible for the poor CME production of AR 12192.
\end{abstract}


%\keywords{globular clusters: general --- globular clusters: individual(NGC 6397,
%NGC 6624, NGC 7078, Terzan 8}



\section{Introduction}
As the main sources of solar intense eruptions, including filament eruptions, major flares, and coronal mass ejections (CMEs), active regions (ARs) have been extensively explored \citep[e.g.,][]{hagyard84, wang94, zhang12}.
% since the early stage of the last century \citep{newton30}.
%M- or X-class
%zirin73, deng01, su14
Strong flares tend to occur near the prime magnetic polarity inversion lines (PILs) of ARs where the field gradients are steep and the horizontal components are highly sheared \citep[e.g.,][]{wang02, deng06, van09}.
%wangh94
In many previous works \citep[e.g.,][]{wang94, schrijver05, zhang07, sun12}, shear and emergence of magnetic flux were suggested to be an important way to store the non-potential energy in the coronal field of ARs, which might be released in the subsequent eruption.
%wang04
%low77, moon02

%A variety of models have been devoted to interpreting the eruption initiations \citep[as reviewed by, e.g.,][]{lin03,vrsnak08, chen11}.
Magnetic reconnection in destabilization processes is considered as a possible mechanism to trigger solar eruptions \citep[][]{vrsnak08}.
%lin03, vrsnak08
In the tether-cutting (TC) model \citep{moore01}, 
%or a similar flux-cancelation mechanism discussed by \citet{van89}, \citet{moore80} and 
the reconnection between the inner legs of the sheared core field would lead to an unbalanced situation between the upward magnetic pressure force and downward magnetic tension force. This may drive the reconnected field connecting the far ends of the core field and the envelop field to expand outward.
Then, another reconnection occurs below the expanding field and further speeds up the eruption to form the CME \citep[e.g.,][]{amari14, chen14}.
Sometimes, the inflating field stops when reaching a certain height and no obvious CME would be observed in the wake of the eruption \citep[e.g.,][]{ji03}. 
The related flares are called confined flares \citep[e.g.,][]{wang07}.
% \citep[e.g.,][]{guo10, cheng11, yang14}. 
%%liu09
%and \textbf{Extreme Ultraviolet (EUV)} wave 

Strong confinement from the overlying magnetic arcades is believed to play an important role in the failed eruptions \citep[e.g.,][]{torok05, guo10, shen11, netzel12, chen13}.
To characterize how fast the overlying magnetic field decays, a decay index is usually defined as $n$ = --$d$ log($B_{h}$)/$d$ log($H$).
Here, $B_{h}$ denotes the horizontal component of the potential field strength; $H$ is the height above the solar surface.
In some earlier studies \citep{bateman78, kliem06, fan07, aulanier10}, a typical threshold of $n$ for torus instability initiation was suggested to be in the range [1.5, 2.0].
%In the theoretical \citep{bateman78} and simulation \citep{aulanier10} studies, $n$ = 1.5 was suggested to be a critical value for torus instability \citep[][]{kliem06} initiation.
When $n$ reaches a larger value at the height of the erupting flux rope, torus instability would lead to a full eruption of the system.
% \citep{kliem06}.
%Some other works \citep[e.g.,][]{fan07, liu08, demoulin10, xu12} extended the related investigations.

AR 12192 recently attracted considerable attention \citep[e.g.,][]{thalmann15, sun15} for its large sunspot group and high flare productivity.
According to the $GOES$ observations, while AR 12192 passed across the visible solar disk, it produced six X-class and twenty-nine M-class flares from October 18 to 29.
%, and more than sixty-three C-class
However, it is surprising that only one M-flare was associated with a CME.
The data from the Atmospheric Imaging Assembly/$Solar\ Dynamic\ Observatory$ \citep[AIA/$SDO$;][]{lemen12} show that some of the flares had similar origin within the AR and common spatial and timing characters, implying they were homologous flares with analogous triggering mechanism \citep[e.g.,][]{sui04, yang14}.
%and heating process
To probe the initiation mechanisms of these homologous flares and causes of the low CME-association of the major flares, we examined the evolution of the photospheric magnetic fluxes, the major flares (M- and X-class), and the background potential field of this AR. 
%rare correlation between the major flares and CMEs

%Why and how did these flares occur? Why were the flares rarely correlated with CMEs?
%To answer these questions, 

%A critical decay index was suggested theoretically to range from 1.5 to 2.0 (Kliem \& To?ro?k 2006), while MHD simulations gave values of 1.53 (To?ro?k \& Kliem 2005) and 1.9 (Fan \& Gibson 2007) for a full eruption of a flux rope.
%To quantitatively describe how fast the strapping field decays, a decay index is defined as n = ?d log(Bt )/d log(h) (Kliem 2006), in which Bt is the strength of the strapping field in the transverse direction and h is the radial height above the photosphere. 

%\begin{equation}
%n=\sqrt{\textup{EM}/l}
%\end{equation}

%At times, the filament eruptions are ``failed'' \citep[e.g.,][]{ji03,liu09,mrozek11,netzel12,chen13}.

%since the existence of magnetic fields in sunspots was discovered (Hale 1908??).
%Many researches were focus on the 

%The Atmospheric Imaging Assembly (AIA; Lemen et al. 2012) on the Solar Dynamics Observa- tory (SDO) images the solar atmosphere in seven EUV nar- row wavebands, covering a wide temperature range of from 0.05 MK (304 , Heii) to 20 MK (193 , Fexxiv). This broad temperature coverage allows for a fuller investigation on the overlying magnetic structures in failed filament eruptions. 

%The remainder of the paper is organized as follows. In the next section, we briefly describe the observations and data used in our study. This is followed by a detailed study of the filament eruption and the overlying EUV arcades. Finally, we give the conclusions and discussions.

\section{Data and Observations}
%\usepackage{enumerate}
%\begin{enumerate}[label=\emph{\roman{*}})]
%\item 
%We mainly used the data from AIA/SDO.
%The Atmospheric Imaging Assembly/$Solar\ Dynamic\ Observatory$ \citep[AIA/$SDO$;][]{lemen12} 
The Helioseismic and Magnetic Imager \citep[HMI;][]{schou12} on $\it{SDO}$ produces photospheric vector magnetograms with $\sim$0.$\arcsec$5 plate scale at 12 minutes cadence \citep{hoe14}.
In the present study, we used de-projected maps in cylindrical equal area coordinates of the automatically identified AR \citep{bobra14}.
%We used the data product from HMI, called Space-weather HMI Active Region Patches (SHARPs), which provides de-projected maps in cylindrical equal area coordinate 
The vertical magnetic field component (B$_{z}$) was utilized for the flux estimation and as the lower boundary condition for potential field modeling.
%As noted by \citet{sun15}, 
The HMI B$_{z}$ data show significant errors, especially in the negative-polarity umbra of the following sunspot. We used an interpolation method to replace the defective data with fitting values. 
%centered at the data point with the minimum value and performed a linear fitting to the good data selected by the empirical conditions along different radius. Then, the ``bad'' data were replaced by the fitting results. 
According to our calculations, the maximum difference between the fluxes from the original and corrected data is only $\sim$2\%, which does not affect our conclusions.
%We used the photospheric vector magnetic field data \textbf{with a spatial resolution of $\sim$1.$\arcsec$0 \citep{hoe14}}  and 12 minutes cadence from the Helioseismic and Magnetic Imager \citep[HMI;][]{schou12} on $\it{SDO}$ to analyze the photospheric evolution of AR 12192.
%The AIA provides full-disk images up to 0.5 R$_{\sun}$  above the solar limb with 0.$\arcsec$6 pixel size. 
%We mainly used 304 \AA, 1600 \AA, 171 \AA, and 94 \AA\ images.
%We mainly used the data (Level 1.5 images) at four channels \textbf{peaked} at 304 \AA\ (\ion{He}{2}, 0.05 MK), 1600 \AA\ (\ion{C}{4} + continuum, 0.1 MK), 171 \AA\ (\ion{Fe}{9}, 0.6 MK), and 94 \AA\ (\ion{Fe}{18}, 7 MK).
%, and 131 \AA\ (\ion{Fe}{8}, 0.4 MK and \ion{Fe}{21}, 11 MK).   1.$\arcsec$008
To better show the similarities of the homologous flares, we de-rotated the AIA data for four X-flares (XF2--XF5; see Table~1) to the same time (October 23 15:00 UT).
% and the connectivities of the sheared arcades in the AR.
%the Helioseismic and Magnetic Imager \citep[HMI;][]{schou12} on $\it{SDO}$
%\end{enumerate}
%\section{Results}
\section{Results}
\subsection{Statistics of the Major Flares} \label{}
The related information of the six X-flares (XF1--XF6) and twenty-nine M-flares (MF1--MF29) are listed in Table~1.
The mean duration (from the start to end time based on $GOES$ SXR event list\footnote[1]{http://hesperia.gsfc.nasa.gov/goes/goes$_{-}$event$_{-}$listings/}) of the X-flares is $\sim$69 minutes, which is longer than that ($\sim$32 minutes) of the M-flares.
%goes_xray_event_list_2014.txt,  
%However, the longest duration is from the M-class flare MF24, which seemed to include about three energy release processes.
By checking the AIA data, we find that not all the flares originated from the AR core. 
Five M-flares appeared at two locations (L1 and L2; see Figures 1(a) and (d)) on the AR periphery. MF4 and MF7 occurred at L1; MF11, MF20, and MF25 took place at L2.
%Almost all the flares were not accompanied with any large-scale plasma \textbf{and} magnetic structure ejection.
Among the thirty-five flares, only MF11 produced a CME.
Jet activities were observed in the four periphery flares MF4, MF7, MF11, and MF20.
%to be associated with

%a angular width of $\sim$82$\degr$ and
%According to $\it{GOES}$-15 observations, a C5.1 flare took place in NOAA AR 11990 (S12E82) from 21:31 UT on 2014 Feb 24.
%The AIA data show that no mass or magnetic structure escaped from the solar surface during the weak flare. 
%Three hours later, an X4.9 flare occurred from the same location with peak at 00:49 UT on Feb 25, accompanied by a filament eruption and a halo CME with a median velocity of $\sim$1041 km s$^{-1}$ (see CACTus catalogue, http://sidc.be/cactus).
%Our observations cover both events well.

\subsection{Photospheric Magnetic Field Evolution}
The HMI B$_{z}$ maps in Figures~1(a)--(f) present the general evolution of AR 12192 from October 21 to 26.
We notice that three pairs of magnetic fluxes (labeled with ``N$_{1}$P$_{1}$'', ``N$_{2}$P$_{2}$'', and ``N$_{3}$P$_{3}$'' in Figure~1) appeared in the AR core.
In the AIA 94 \AA\ image (see Figure~3(d)), it can be seen that some coronal loops connected N$_{1}$ (N$_{3}$) and P$_{1}$ (P$_{3}$).
%Owing to lower altitude, the loops between N$_{2}$ and P$_{2}$ seem to be occulted by the overlying arcades. 
The HMI vector magnetic field data (Figure~1(g)) also reveals the connectivity between N$_{2}$ and P$_{2}$.
%that N$_{2}$ and P$_{2}$ were coupled with each other
Figure~1 shows that the negative and positive fluxes of N$_{1}$P$_{1}$ and N$_{3}$P$_{3}$ were respectively located at two sides of the AR prime PIL.
As the AR developed, P$_{1}$ and P$_{3}$ moved southwest  along the direction 
%(indicated by the thick blue arrows in Figures~1(a) and (b))   the positive fluxes 
approximately parallel to the AR prime PIL and gradually decayed.
It is apparent that this kind of shearing motions of photospheric fluxes would result in the shear formation and strengthening of the coronal field in the AR.
%From the HMI vector magnetogram (Figure~1(g)), 
In Figure~1(g), we also see that the horizontal field in the AR core were basically aligned with the AR prime PIL.
Different from N$_{1}$P$_{1}$ and N$_{3}$P$_{3}$, N$_{2}$P$_{2}$ appeared near the prime following sunspot of the AR.
During its emergence, it collided and partly canceled out with the ambient magnetic field.
%Two locations L1 and L2 at the AR periphery are indicated by the yellow circles in panels (a) and (d), respectively. 
%, which were respectively associated with MF4, MF7, MF11, and MF20.

% (animation~1 in the online journal)
%During its disk passage, we noticed three pairs of magnetic fluxes, which are respectively labeled with ``N$_{1}$P$_{1}$'', ``N$_{2}$P$_{2}$'', and ``N$_{3}$P$_{3}$'' in Figure~1, in the core region of the AR.
%\begin{equation}
%n=\sqrt{\textup{EM}/l}
%\end{equation}
\subsection{Jets Associated with the Flares}
%As mentioned above, 
Four jets occurred from L1 and L2.
Using the AIA 304 \AA\ filtergrams, we display the four jets and their accompanying flares in Figures~2(a)--(c) (MF4, MF7, and MF20, respectively) and Figures~3(a)--(c) (MF11).
In the AIA 171 \AA\ channel (see animation~2), both bright and dark \textbf{features} were \textbf{observed} along the open field or the legs of large-scale coronal loops to form the jets.
%during the flares.
%According to our calculations, 
The maximum projected lengths of the four jets are $\sim$282, $\sim$281, $\sim$130, and $>$569 Mm (beyond the FOV of AIA), respectively; their respective mean velocities (the maximum projected lengths divided by the corresponding durations) are 361$\pm$1, 360$\pm$3, 309$\pm$6, and 379$\pm$3 km s$^{-1}$; and they separately have lifetimes of $\sim$46, $\sim$61, $\sim$74, and $\sim$172 minutes.
Compared to the surges or jets studied by \citet{chen08}, which also occurred at the AR border but with microflares, the speeds, spatial scales, and lifetimes found here are evidently greater.
The HMI B$_{z}$ data clearly show that the magnetic fluxes of different polarities emerged, converged, and canceled with each other at the base of the jets before the associated flares. Similar results were also found in earlier studies \citep[e.g.,][]{jiang07, chen08}.

MF11 is the only one followed by a CME,
%\textbf{According to the CACTus catalog\footnote[2]{http://sidc.be/cactus}, the median velocity of the CME is $\sim$496 km s$^{-1}$.}
for which $CACTus$\footnote[2]{http://sidc.be/cactus} lists a median velocity of $\sim$496 km s$^{-1}$.
From Figure~3(a), we can see that a small arch filament existed at the flare source region before the onset of MF11.
%From Figure~3(a), we can see that 
As the flare occurred, the hot plasma was ejected southwest at first (as shown by the curved arrows in Figures~3(b) and (c)) and then spurted out southeast more radially (as indicated by the straight arrow in Figure~3(c)).
The 171 \AA\ filtergrams (see animation~4) clearly reveal that the ejected plasma moved along one leg of a large-scale loop at the beginning and then turned to the ambient open field.
According to our measurement, the \textbf{projected} deflection angle of the ejection direction is $\sim$40$\degr$.
After MF11, the arch filament disappeared, indicating its participation in the ejection.
According to \citet{moore10}, this jet is very likely a blowout jet, in which the small arch filament underwent a miniature version of a blowout eruption and produced the subsequent CME \citep[e.g.,][]{hong11, pucci13}.

\subsection{Tether-cutting Reconnection Triggering the Homologous Confined Flares}
%Thirty flares under our investigation, including six X- and twenty-four M-flares, arose from the core region of AR 12192.
\textbf{Among the thirty flares from the AR core, we mainly focused on the four on-disk X-flares (XF2--XF5). These flares} exhibited similar flaring structures (see the animation~3), indicating they are likely homologous flares.
% \citep[][]{sui04}.
%with analogous triggering mechanism and heating process
Coronal emission associated to three of the homologous flares (XF2--XF4) are shown in Figures~2(d)--(f) and Figures~2(g)--(i), respectively.
The AIA 1600 \AA\ images clearly display their two-ribbon structure (FR1 and FR2), which show great similarities concerning spatial distribution and morphology.
In comparison with XF2, the northeastern ends of FR1 in XF3 and XF4 turned to north to extend, 
%(indicated by the arrows in Figures~2(e) and (f)), 
which might be related to the newly emerged N$_{2}$P$_{2}$.
In the 94 \AA\ images, sensing hot flare plasma at a characteristic temperature of 6.3 MK, bright (flare) loops (their apparent projected shape indicated by the dotted lines in Figures 2(g)-(i)) connecting FR1 and FR2 suggest a complex, non-potential field geometry, which are different from the usual post-flare loops in the eruptive flares.
%and 131 \AA\
This implies that only a little portion of the non-potential energy was released during these confined flares, which is consistent with the model results reported by \citet{thalmann15} and \citet{sun15}.
 %In Figure~1(g), the outlines of the two ribbons in XF3 are overlaid on the HMI vector magnetogram. 
 
It is very likely that TC reconnection triggered these homologous flares. 
As an example, we display the evolution of XF3 in Figures~3(d)--(f) and 
Figures~3(g)--(i), which correspond to the 94 \AA\ and 304 \AA\ images respectively.
In 94 \AA, it can be seen that some sheared magnetic arcades existed in the AR core prior to the flare.
%The formation of these sheared loops should have a close association with the shearing motions of the photospheric magnetic fluxes described previously??.
We outline two sheared loop systems (``AB'' and ``CD'') in Figure~3(d).
%, which are respectively shown by the yellow and red dotted lines.
%The locations of the ends of AB and CD in the photosphere are marked in Figure~1(d).
Figure~1(d) shows the locations of their ends in the photosphere.
It is evident that AB and CD straddled the AR prime PIL and connected the opposite-polarity field on the two sides.
As the flare started and developed, AB reconnected with CD, which led to the formations of the small flaring loop BC (Figure~3(e)) and large-scale loop structure AD (Figure~3(f)). 
Meanwhile, a jet along CB was also observed not only in the hot 94 \AA\ line (Figure~3(e)) but also in the cool 304 \AA\ channel (Figure~3(h)), which provides the evidence of magnetic reconnection.
%Meanwhile, the hot mass flow from C to B were also observed not only in the hot 94 \AA\ line (see Figure~3(e)) but also in the cool 304 \AA\ channel (see Figure~3(h)), which may be caused by the chromospheric evaporation during the flare.
Figure~3(f) shows that the post-flare loops still retained a highly-sheared topological structure, which made possible the next homologous energy release.
In Figure~3(g), some filament fibrils were almost aligned with the AR PIL before the flare. However, the detailed 304 \AA\ data reveal that these fibrils were not affected by the eruption, suggesting a relatively-high reconnection site in the flare.
This is in agreement with what was suggested by \citet{thalmann15}, based on the large initial separation of  the associated flare ribbons.
Our results strongly support the TC reconnection mechanism. In addition, the newly emerging flux tubes reconnecting with pre-existing coronal loops may also play a role as the enhanced emission of the loops nearby N$_{2}$P$_{2}$ was detected before the flare (see animation~5).

%\textbf{Although our results strongly support the TC reconnection mechanism, we cannot entirely exclude other possibilities. For example, the enhanced emission of the loops nearby N$_{2}$P$_{2}$ observed before the flares indicates newly emerging flux tubes reconnecting with pre-existing coronal loops likely play a certain role in these eruptions.}
%\textbf{In addition, before these X-flares started, we also observed the enhanced emission of the loops nearby N$_{2}$P$_{2}$, which might involve a different reconnection scenario, e.g., newly emerging flux tubes reconnecting with pre-existing coronal loops.}

\subsection{Strong Confinement from the Overlying Background Field}
One outstanding feature of AR 12192 is its poor CME production rate, despite the many major flares observed during its disk passage.
According to the previous studies \citep[e.g.,][]{fan07}, the confinement from the overlying background field of AR may play an important role in these confined flares.
Utilizing the HMI B$_{z}$ maps, the potential field extrapolation of AR 12192 is derived from the Green's function method \citep{chiu77, metcalf08}.
We calculate the decay index $n$ of the horizontal extrapolated field B$_{h}$ along a reference line (see Figure~1(c)), the orientation of which is basically aligned with the AR prime PIL and changes with the AR evolution.
To determine the reference line {in a certain B$_{z}$ map, we firstly obtain the positions of the maximum positive and negative fluxes.
%\textbf{use the corresponding B$_{z}$ map to}
%, which are usually at the centers of the prime leading and following sunspots respectively.
%(respectively marked with the blue and yellow pluses in Figure~1(c))
% of positive and negative polarity.
Then, we connect them and plot the reference line with a fixed length through the midpoint. 
The angle we chose between the reference line and connecting line is 60$\degr$.
With this choice, the direction of the AR prime PIL is outlined best in most of the cases.
%The angle between the reference line and connecting line is \textbf{assigned} 60$\degr$\textbf{, with which the direction of the AR prime PIL was outlined best in most of the cases.}

Generally, $n$ varies with the height above the photosphere. 
By checking two limb flare cases (MF1 and XF1), 
%By checking the AIA 131 \AA\ data of MF1 and XF1 (near the solar eastern limb), 
we find that the height ranges of the erupting flare loops are respectively $\sim$47--115 Mm and $\sim$78--110 Mm, comparable to the result of 42--105 Mm suggested by \citet{liu08} and \citet{xu12}.
In this study, we therefore calculate the mean values of $n$ for the height range $\sim$40--105 Mm above the photosphere at every point along the reference line firstly and then average them to derive the mean decay index $\overline{n}$, the time variation of which is displayed in Figure~4(a).
%in the vertical plane above the prime PIL.
%The mean decay index $\overline{n}$ is derived from all of the mean values of n and its time variation is displayed in Figure~4(a).}
%\textbf{The time variation of $\overline{n}$} is displayed in Figure~4(a).
We only calculate $\overline{n}$ from October 20 to 26, during which the longitude of the AR center was within $\sim$E42--W46.
It can be seen that all $\overline{n}$ is always less than 1.5, the lower limit of the typical threshold for torus instability onset \citep{kliem06, liu08}.
Even taking the uncertainties into account, the maximum of $\overline{n}$ is $\sim$1.63 and does not exceed the upper limit (2.0) of the threshold.
%bateman78, aulanier10
%Considering the uncertainty, $\overline{n}$ varies in the range of $\sim$[1.2, 1.6], implying a strong level of the horizontal coronal field \citep[e.g.,][]{?}.
%It can be seen that \textbf{$\overline{n}$ with 1$\sigma$ uncertainty varies between 1.23 $\pm$ 0.07 and 1.50 $\pm$ 0.11, which are smaller than those ?? and implies a strong level of the horizontal coronal field \citep[e.g.,][]{liu08, xu12}.?}
From October 20 to 22, $\overline{n}$ increases from $\sim$1.41 to $\sim$1.50 firstly and then keeps stable for about 20 hr.
After October 22 14:00 UT, $\overline{n}$ begins to decline and finally reaches a value of $\sim$1.23 at the end of October 26.
We point out that different from previous studies \citep[e.g.,][]{liu08, xu12, sun15}, we firstly calculate and analyze the time evolution of the mean decay index of an AR horizontal background field during a long period (7 days).

Figure~4(b) displays the time profiles of the positive (red pluses) and unsigned negative (blue pluses) vertical fluxes of the whole AR during the same period.
%To avoid the possible instrumental errors, we have smoothed the HMI B$_{z}$ data within 10 hr.
From October 20 to 26, both the positive and unsigned negative fluxes decrease at first and then are enhanced gradually, which seems to have an inversed time evolution of $\overline{n}$.
Since the background potential field of AR might be strengthened (weakened) with the emergence (cancelation) of the photospheric fluxes, this result seems reasonable.
The similar magnitude of the total positive and total unsigned negative flux also suggests a relatively close coronal field structure of the AR.
%11\%
%According to our results, to some extent, the mean \textbf{decay} index  $\overline{n}$ increases with the decline of magnetic fluxes and decreases with the emergence of the fluxes. 

%the horizontal field strength
Figures~4(c)--(f) show the height profiles of B$_{h}$ about 1 hr before the start (blue dashed line) and 1 hr after the end (red dotted line) of XF2--XF5, respectively.
The decay indices $n$ (averaged along the reference line) derived from B$_{h}$ are displayed by the green dashed (pre-flare) and purple dotted (post-flare) curves in the four panels.
%\textbf{The decay indices $n$ (averaged along the reference line) derived from the preflare (postflare) \textbf{B$_{h}$} are displayed by the green dashed lines (purple dotted lines) in the four panels.}
%The decay indices $n$ (averaged along the reference line) derived from the preflare \textbf{(green dashed line) and postflare (purple dotted line)} horizontal field strengths are displayed in the four panels as well.
The strength and height are in logarithm units.
It is apparent that there are little changes of B$_{h}$ and $n$ during the flares.
%the horizontal potential field
Correspondingly, the mean decay indices $\overline{n}$ before and after the flares are also similar (see the green and purple values of $\overline{n}$ in Figures~4(c)--(f)).
According to our results, $\overline{n}$ are $\sim$1.47$\pm$0.14, $\sim$1.37$\pm$0.09, $\sim$1.32$\pm$0.07, and $\sim$1.27$\pm$0.07 for the four confined X-flares XF2--XF5, respectively.
Within the uncertainty ranges, our results agree with those of \citet{sun15}.

\section{Summary and Discussion}
We made a statistics of six X-flares and twenty-nine M-flares in AR 12192 from October 18 to 29 and performed a detailed investigation to four of the X-flares and four of the M-flares.
%We \textbf{investigated} six X-class and twenty-nine M-class flares in AR 12192 from October 18 to 29.
%We conducted a detailed investigation of six X-class and twenty-nine M-class flares in AR 12192 from October 18 to 29.
Our main results are as follows:
%\begin{enumerate}[label=\emph{\roman{*}})]
%\renewcommand{\theenumi}{\Roman{enumi}}
%\renewcommand{\labelenumi}{(\theenumi)}
\begin{enumerate}
%\item[(1)]
\item 
From October 20 to 26, the background field of AR 12192 remained at a strong level. 
The mean decay index $\overline{n}$ of the horizontal potential field in a vertical plane aligned with the prime PIL and within the height range of 40--105 Mm varied in the range [1.23, 1.50], implying that a strong confinement from the overlying magnetic field may play an important role in the poor CME production of the AR.
This is in agreement with the findings of \citet{sun15} and \citet{thalmann15} for the individual cases XF3 and XF2, respectively.

\item Four periphery M-flares were accompanied by jets. The emergences and cancelations of the photospheric fluxes with different magnetic polarities were observed at the roots of the jets, supporting the standard model for solar coronal jets \citep[e.g.,][]{moore10, chen12, pariat15}.

\item 
In HMI B$_{z}$ maps, the apparent shearing motions of emerged photospheric fluxes were found in the AR core.
This might have established a coronal field configuration suitable for TC reconnection, which was observed in four homologous confined X-flares.
These observations are well consistent with the TC model described in \citet{moore01}.
\end{enumerate}

TC reconnection is a possible way to trigger solar eruptions \citep[e.g.,][]{liu10, sterling11, amari14}. 
\citet{chen14} exhibited a detailed process of TC reconnection during a major event, which was followed by a halo CME.
However, as reported here, CME would not always appear in the eruptions triggered by TC reconnection \citep{aulanier10}.
Several similar observational cases with weak flares have been discussed in \citet{moore01} and \citet{chen14}.
To our knowledge, we firstly report that successive strong X-flares were triggered by TC reconnection, but not yielded any partial or full eruptions.

%In the magnetohydrodynamic (MHD) simulation of \citet{aulanier10}, it has been found that TC coronal reconnection do not trigger CMEs in bipolar magnetic fields, but can build up and raise flux ropes to the critical height at which the torus instability causes the final eruption.
%Our observations coincide well with their simulation results.

%Different from the previous studies \citep[e.g.,][]{liu08, xu12, sun15}, we firstly calculated and analyzed the temporal evolution of the decay index of the horizontal background field during a long period (7 days).
%To our knowledge, this is the first related investigation on this aspect.
%It will be helpful in improving the torus instability mechanism of solar eruption to continue similar investigations on more ARs in the future.

%between the sheared loops in AR   above the photosphere 
%We conducted a detailed investigation of the thirty-five flares with GOES level above M-class in AR 12192 from October 18 to 29.
%of different magnetic polarities
\acknowledgments
%The authors sincerely thank the referee for very helpful and constructive comments that improved this paper.
%We acknowledge the AIA team for the easy access to calibrated data. The AIA data are courtesy of SDO (NASA) and the AIA consortium.
%$SDO$ is a mission of NASA's Living With a Star Program.
%The STEREO/SECCHI data are produced by an international consortium: NRL, LMSAL, NASA, GSFC (USA); RAL (UK); MPS (Germany); CSL (Belgium); and IOTA, IAS (France). We also thank the LASCO/SOHO team for the data support.
The $SDO$ data are courtesy of NASA, the $SDO$/HMI and AIA science teams.
This work was supported by NSFC (41204124, 11221063, 11203037, and 41331068), the Project Funded by China Postdoctoral Science Foundation (2015M571126), the National Basic Research Program of China under grant G2011CB811403, and the Strategic Priority Research Program---The Emergence of Cosmological Structures of the Chinese Academy of Sciences (No. XDB09000000).
%National Natural Science Foundation of China
%the CAS project KJCX2-EW-T07
%S.M. thanks Kelly Korreck, Mark Webb, Ed Deluca, and MeredithWills for helpful discussions. S.M. is also grateful to Steven Cranmer,Alexander Engell, and HenryWinter for help with techniques. At SAO, this work was supported by subcontract SP02H1701R from Lockheed Martin, by NASA contract NNM07AB07C, and by NASA grant NNX08BA97G. J.C.R. acknowledges support from NASA grant NNX11AB61G. J.L.¡¯s work was supported by Program 973 grant 2011CB811403 and NSFC grant 10873030.

%\bibliographystyle{apj}
%\bibliography{ms_1103.bib}
%\begin{thebibliography}{}
%\bibitem[Priest \& Forbes(2002)]{priest02} Priest, E. R., \&   Forbes, T. G., 2002, \aapr,10, 313
%\end{thebibliography}

\clearpage

\clearpage

\begin{deluxetable}{cccccccccr}
\tabletypesize{\scriptsize} % 8pt
%\tabletypesize{\small}   % 11pt
%\tabletypesize{\footnotesize} % 10pt
%\rotate
\tablecaption{X- and M-class flares in Solar AR 12192 \label{tbl-1}}
\tablewidth{0pt}
\tablehead{
\colhead{Event} & \colhead{Date} & \colhead{Start Time} & \colhead{Peak Time} & 
\colhead{Duration}  & \colhead{$\it{GOES}$} &
\colhead{Location} & \colhead{CME} & \colhead{Other}\\
\multicolumn{1}{c}{} & \multicolumn{1}{c}{} & \multicolumn{1}{c}{(UT)} & \multicolumn{1}{c}{(UT)} & \multicolumn{1}{c}{(minutes)}  & \multicolumn{1}{c}{Level} & \multicolumn{1}{c}{in AR} & \multicolumn{1}{c}{Production} & \multicolumn{1}{c}{Activities}
%\colhead{Event} & \colhead{Date}\
%\colhead{}    & \colhead{}    &  \colhead{Start} &  \colhead{Peak} \\
%\cline{3-5}
%\colhead{}    &  \multicolumn{3}{c}{Non-shell Stars} &   \colhead{}   &
%\multicolumn{3}{c}{Shell Stars} \\
%\cline{2-4} \cline{6-8} \\
}
\startdata
%\cutinhead{X-class flares}
\sidehead{X-class Flares:}
XF1 & 19-Oct-2014 & 04:17 & 05:03 &  91  & X1.1 & core &
No & \\
XF2 & 22-Oct-2014 & 14:02 & 14:28 &  48  & X1.6 & core &
No &\\
XF3 & 24-Oct-2014 & 21:07 & 21:41  &  66  & X3.1 & core &
No &\\
XF4 & 25-Oct-2014 & 16:55  & 17:08  &  76  & X1.0 & core &
No &\\
XF5 & 26-Oct-2014 & 10:04  & 10:56  &  74   & X2.0 & core &
No &\\
XF6 & 27-Oct-2014 & 14:12  & 14:47  &  57  & X2.0 & core &
No &\\
%\cutinhead{M-class flares}
\sidehead{M-class Flares:}
MF1 & 18-Oct-2014 & 07:02  & 07:58  &  107   & M1.6 & core &
No &\\
MF2 & 20-Oct-2014 & 09:00 & 09:11   &  20   & M3.9 & core &
No &\\
MF3 & 20-Oct-2014 & 16:00  & 16:37   &  55   & M4.5 & core &
No &\\
MF4 & 20-Oct-2014 & 18:55  & 19:02   &  9   & M1.4 & periphery (L1\tablenotemark{a})  &
No & jet\\
MF5 & 20-Oct-2014 & 19:53   & 20:04   &  20   & M1.7 & core &
No &\\
MF6 & 20-Oct-2014 & 22:43   & 22:55   &  30   & M1.2 & core &
No &\\
MF7 & 21-Oct-2014 & 13:35   & 13:38   &  5   & M1.2 & periphery (L1) &
No & jet\\
MF8 & 22-Oct-2014 & 01:16   & 01:59   &  72  & M8.7 & core &
No &\\
MF9 & 22-Oct-2014 & 05:11   & 05:17  &  10 & M2.7 & core &
No &\\

MF10 & 23-Oct-2014 & 09:44   & 09:50   &  12  & M1.1 & core &
No &\\
MF11 & 24-Oct-2014 & 07:37   & 07:48   &  16   & M4.0 & periphery (L2\tablenotemark{b}) &
Yes & jet\\
MF12 & 26-Oct-2014 & 17:08   & 17:17   &  22   & M1.0 & core &
No &\\
MF13 & 26-Oct-2014 & 18:07   & 18:15   &  13   & M4.2 & core &
No &\\
MF14 & 26-Oct-2014 & 18:43   & 18:49   &  13   & M1.9 & core &
No &\\
MF15 & 26-Oct-2014 & 19:59   & 20:21   &  46   & M2.4 & core &
No &\\
MF16 & 27-Oct-2014 & 00:06   & 00:34   &  38   & M7.1 & core &
No &\\
MF17 & 27-Oct-2014 & 01:44   & 02:02  &  27   & M1.0 & core &
No &\\
MF18 & 27-Oct-2014 & 03:35   & 03:41   &  13   & M1.3 & core &
No &\\
MF19 & 27-Oct-2014 & 09:59  & 10:09   &  27   & M6.7 & core &
No &\\

MF20 & 27-Oct-2014 & 17:33   & 17:40   &  14   & M1.4 & periphery (L2) &
No & jet\\
MF21 & 28-Oct-2014 & 02:15   & 02:42   &  53   & M3.4 & core &
No &\\
MF22 & 28-Oct-2014 & 03:23   & 03:32   &  18   & M6.6 & core&
No &\\
MF23 & 28-Oct-2014 & 13:54   & 14:06   &  29   & M1.6 & core &
No &\\
MF24 & 29-Oct-2014 & 06:03   & 08:20   &  169\tablenotemark{c} & M1.0 & core &
No &\\
MF25 & 29-Oct-2014 & 09:54   & 10:01   &  12   & M1.2 & periphery (L2) &
No & \\
MF26 & 29-Oct-2014 & 14:24   & 14:33   &  27   & M1.4 & core &
No &\\
MF27 & 29-Oct-2014 & 16:06   & 16:20   &  27  & M1.0 & core &
No &\\
MF28 & 29-Oct-2014 & 18:47   & 18:50   &  5   & M1.3 & core &
No &\\
MF29 & 29-Oct-2014 & 21:18   & 21:22   &  7  & M2.3 & core &
No &\\
\enddata
%\tablenotemark{a}
%% Text for table notes should follow after the \enddata but before
%% the \end{deluxetable}. Make sure there is at least one \tablenotemark
%% in the table for each \tablenotetext.
%\tablecomments{Table \ref{tbl-1} is published in its entirety in the 
%electronic edition of the {\it Astrophysical Journal}.  A portion is 
%shown here for guidance regarding its form and content.}
\tablenotetext{a}{Marked by the yellow circle in Figure~1(a).}
\tablenotetext{b}{Marked by the yellow circle in Figure~1(d).}
\tablenotetext{c}{The unorthodoxly long duration is due to multiple energy releases.}
\end{deluxetable}

%\clearpage

%% Here we use \plottwo to present two versions of the same figure,
%% one in black and white for print the other in RGB color
%% for online presentation. Note that the caption indicates
%% that a color version of the figure will be available online.
%%
\begin{figure}
\epsscale{0.8}
\plotone{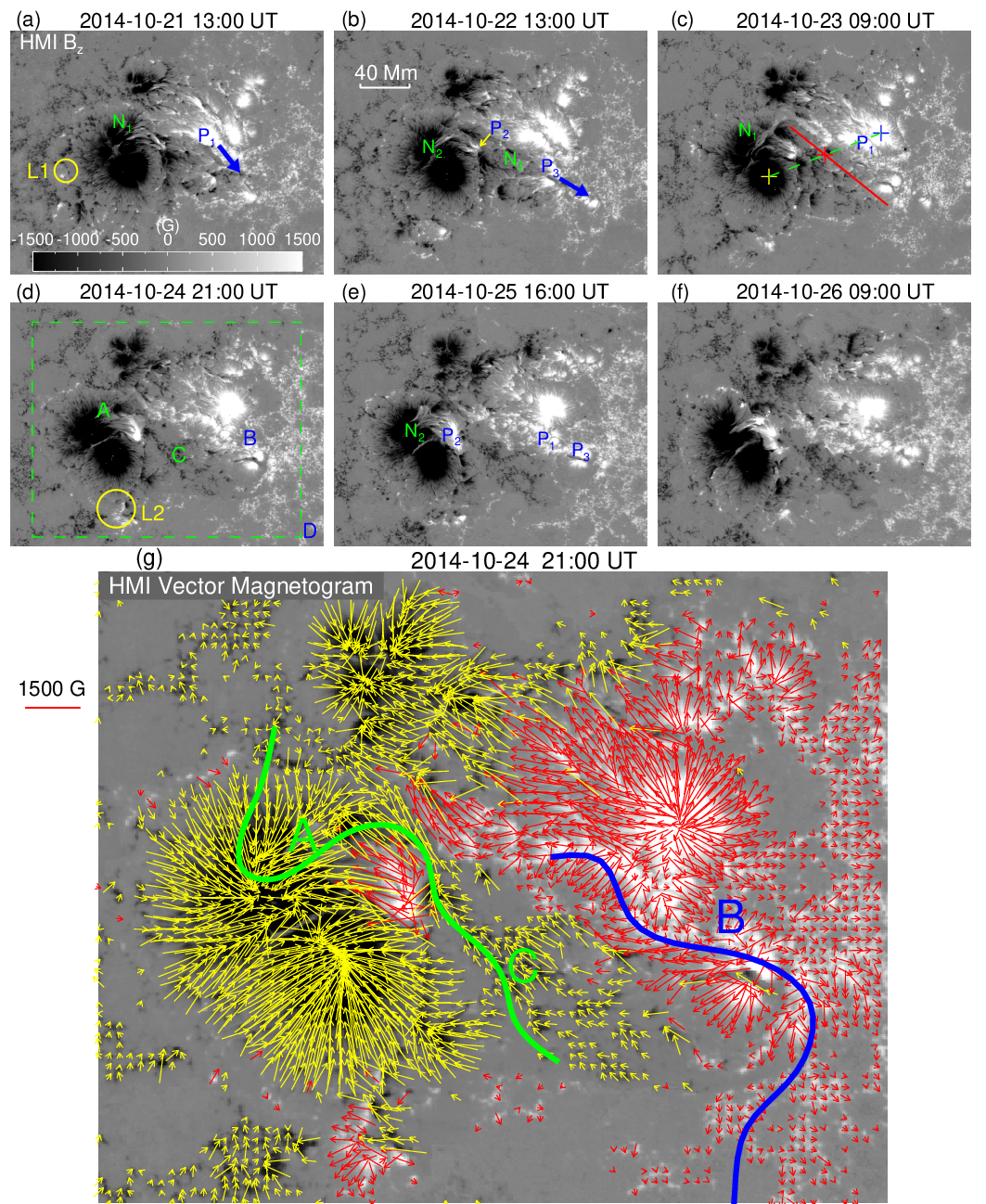}
\caption{((a)--(f)) HMI B$_{z}$ maps (also see the animation~1);
(g) HMI vector magnetic field map. 
The blue thick arrows in panels (a) and (b) indicate the shearing motions of the fluxes P$_{1}$ and P$_{3}$.
The yellow thin arrow in panel (b) points to the location of P$_{2}$.
The yellow, blue and red pluses in panel (c) correspond to the locations of the maximum negative flux, maximum positive flux, and their midpoint, respectively. 
The red line in panel (c) is the reference line along which we calculated the decay index $n$.
The circles in panels (a) and (d) mark the two locations L1 and L2, respectively. 
The dashed box in panel (d) indicates the FOV of panel (g).
The green and blue curves in panel (g) indicate the principal shape of the flare ribbons around the peak time of XF3.
\label{fig1}}
\end{figure}

\begin{figure}
\epsscale{0.9}
\plotone{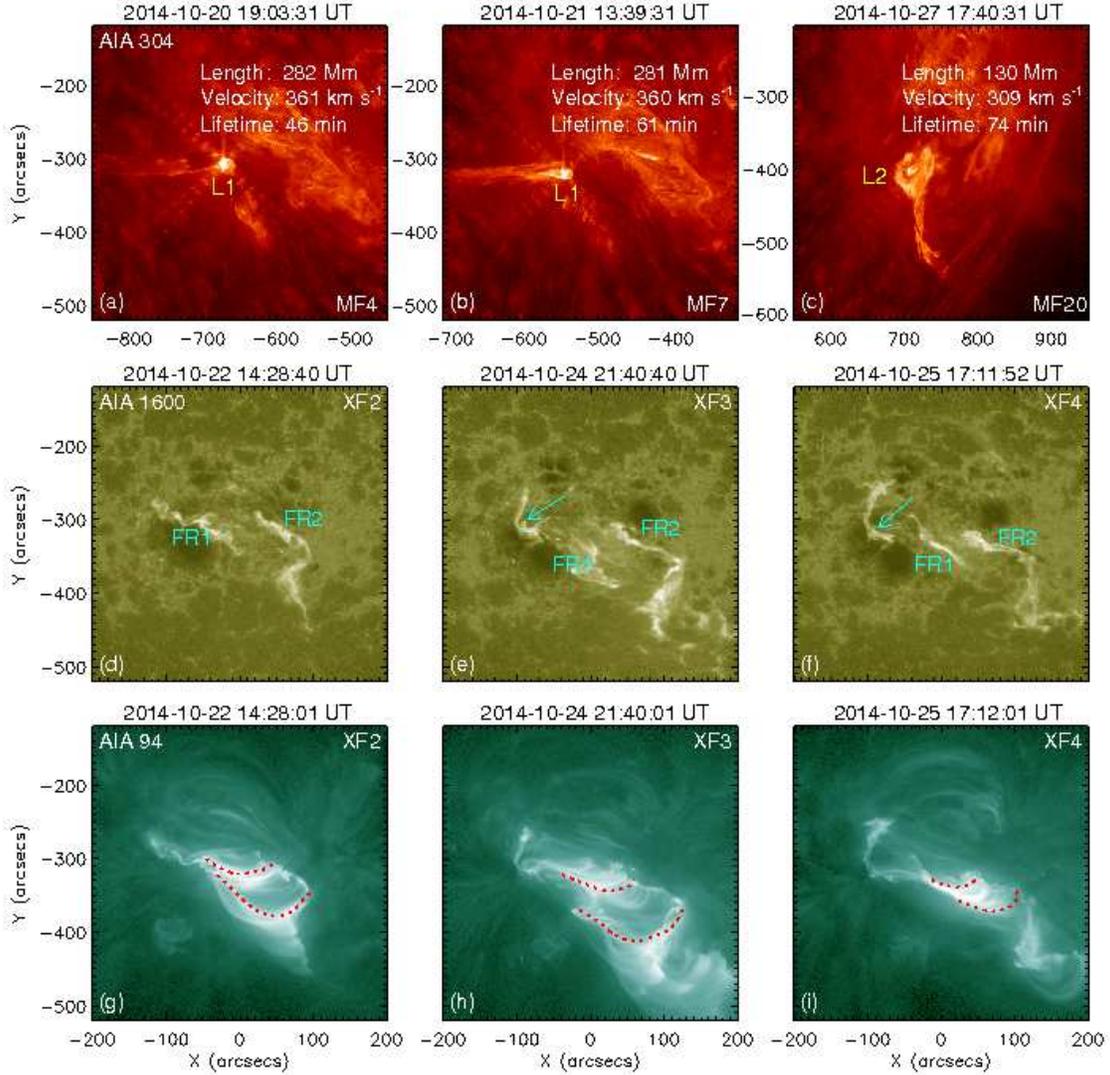}
\caption{((a)--(c)) AIA 304 \AA\ images showing the jets associated with MF4, MF7 and MF11, respectively (also see the animation~2);
((d)--(f)) AIA 1600 \AA\ and ((g)--(i)) AIA 94 \AA\ images displaying the flaring structures of XF2, XF3, and XF4, respectively (also see the animation~3).
%the two flare ribbons and the complex non-potential flaring structures in XF2, XF3, and XF4.
The turquoise arrows in panels (e) and (f) indicate the northward extension of FR1 in XF3 and XF4.
%point to the turning points of FR1 in XF3 and XF4.
%, which may be related to the emergence of N$_{2}$P$_{2}$.
The red dotted curves in panels (g)--(i) outline some apparently sheared and non-potential flare loops during XF2, XF3, and XF4.
%flaring
%AIA 131 \AA\ images showing the eruption of the filament and the associated hot
%flux ropes (also see animation~3).
%The black curve in panel (b) displays the general outline of the hot flux rope formed due to T-C reconnection. 
%The purple and blue dotted lines in panel (c) indicate the two branches of ML$_4$.
%The opened narrow box in panel (d) indicates where the time-slit map in Figure~4(a) was made.
%The FOV is 200\arcsec $\times$ 150\arcsec.
%The FOV is the same as that of Figure~2.
\label{fig2}}
\end{figure}

%\clearpage

\begin{figure}
\epsscale{0.9}
\plotone{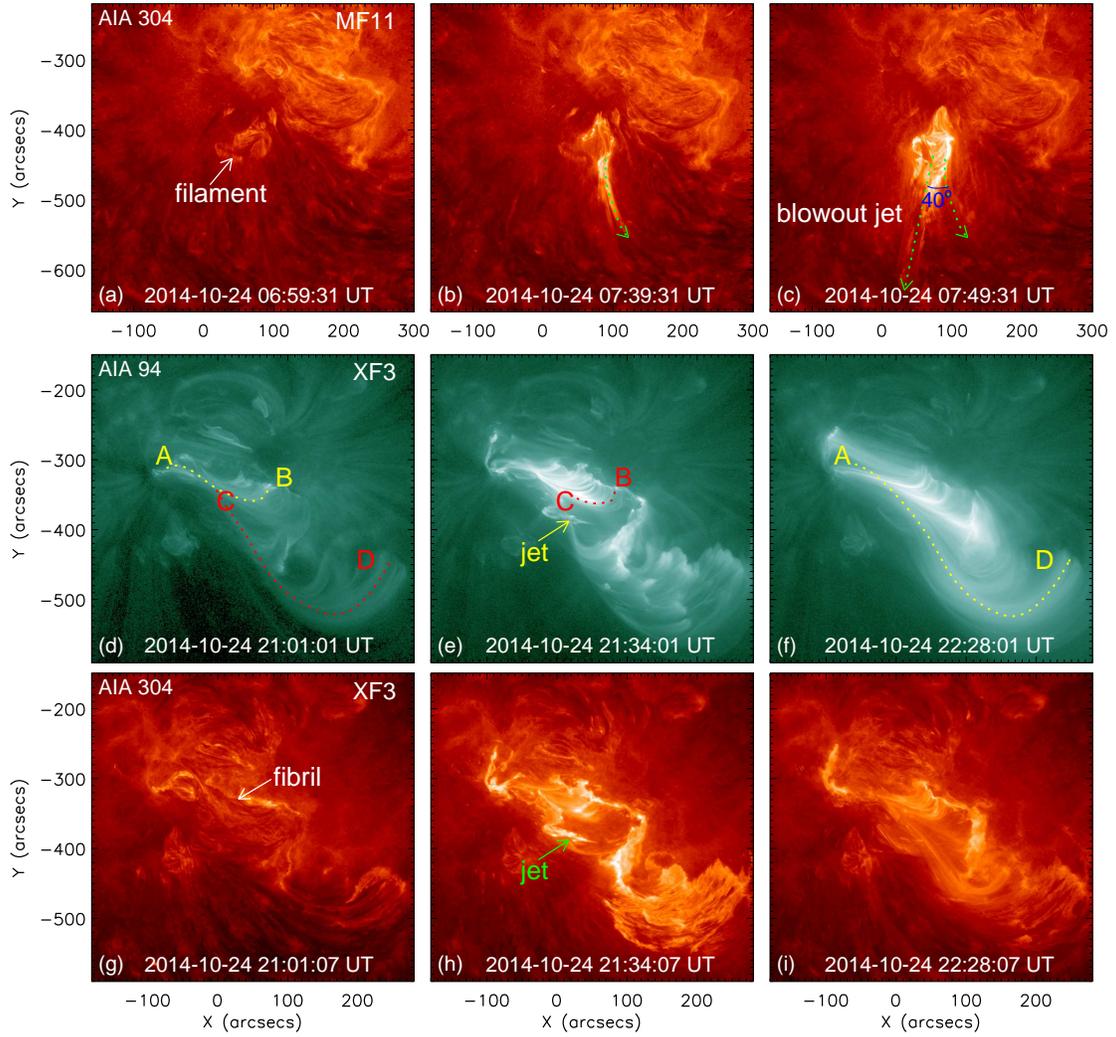}
\caption{((a)--(c)) AIA 304 \AA\ images showing the blowout jet in MF11 (also see the animation~4);
The dotted arrows in panels (b) and (c) indicate the directions of the plasma ejection in the jet.
((d)--(f)) AIA 94 \AA\ and ((g)--(i)) AIA 304 \AA\ images display the evolution of XF3 (also see the animation~5).
%, which was triggered by the TC reconnection between the sheared magnetic arcades ``AB'' and ``CD''.
\label{fig3}}
\end{figure}

\begin{figure}
\epsscale{0.8}
\plotone{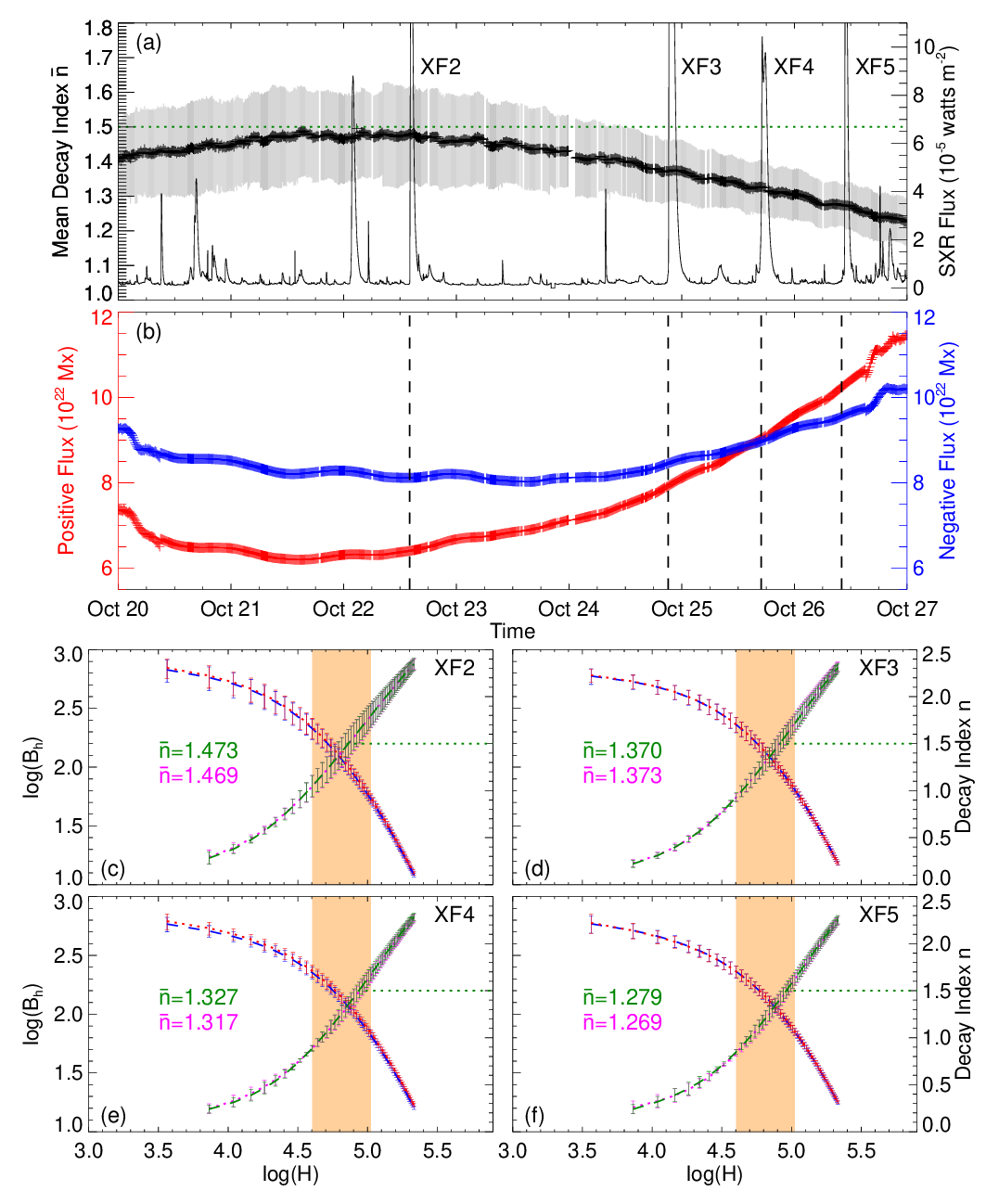}
\caption{(a) Time variations of the mean decay index $\overline{n}$ (plus) with 1$\sigma$ uncertainty (shaded line) and $\it{GOES}$ soft X-ray (SXR) flux (curve);
%of the horizontal potential field
(b) time profiles of the positive (red plus) and unsigned negative (blue plus) vertical fluxes of AR 12192;
((c)--(f)) height profiles of B$_{h}$ with 1$\sigma$ error bars before (blue dashed line) and after (red dotted line) XF2--XF5; height variations of decay index $n$ with 1$\sigma$ error bars before (green solid line)  and after (purple dotted line) the four X-flares.
Horizontal dotted lines in panels (a) and (c)--(f) indicate the lower limit ($n$ = 1.5) of the threshold for torus instability onset.
The vertical lines in panel (b) indicate the start times of XF2--XF5.
The orange regions in panels (c)--(f) mark the height range of 40--105 Mm. 
\label{fig4}}
\end{figure}

\end{document}